\documentclass[prc,twocolumn,floatfix,groupedaddress,nofootinbib,preprintnumbers,amsmath,amssymb,amsfonts,
superscriptaddress,widetable,href]{revtex4}
\usepackage{amsmath,amssymb,amsfonts}
\usepackage{hyperref}
\usepackage{ulem}
\usepackage{graphicx}
\usepackage{color}
\usepackage{longtable}

\newcommand{\be}{\begin{equation}} \newcommand{\ee}{\end{equation}}
\newcommand{\bea}{\begin{eqnarray}} \newcommand{\eea}{\end{eqnarray}}

\allowdisplaybreaks
\begin{document}

\title{\bf  Many-body Dyson equation approach to the seniority model of pairing }

\author{Peter Schuck}
\affiliation{Universit\'e Paris-Saclay, CNRS, IJCLab, IN2P3-CNRS,
91405 Orsay, France\\
Universit\'e Grenoble Alpes, CNRS, LPMMC, 38000 Grenoble, France}

\date{\today}

\begin{abstract}
As is well known, the single level seniority model of pairing has been solved exactly since long using angular momentum algebra. It is shown that it can also be solved using the Dyson equation of standard many-body theory. The formalism shows some interesting many-body aspects.

\end{abstract}

\maketitle

\section{Introduction}

The single level pairing model, known in the literature as the seniority model, see, e.g., \cite{RS} is a cornerstone for the understanding of how pairing works in finite fermion systems, that is in the present case, finite nuclei. The physics of this model was introduced by Racah \cite{Racah} and it has been solved analytically very long ago by  Kerman \cite{Kerman} with angular momentum algebra. To our knowledge, it was never attempted to solve this model purely with many body techniques. The objective of this paper is to show a way how to do this using the Dyson equation together with equation of motion (EOM) techniques. It will turn out that the procedure is not completely trivial revealing at the same time interesting aspects of many-body theory.

\section{The seniority model}

The seniority model is a degenerate single level problem with a constant pairing force.

\begin{equation}
  H=-GS_+S_-
  \label{H}
\end{equation}
with $S_- = \sum_{m>0}c_{-m}c_m~  (S_+ = S_-^{\dag})$ where the summation runs over half the degeneracy of the shell: $\Omega = j+1/2$ with $j$ the spin of the shell, i.e., $\Omega$ is the maximum possible number of pairs $S_+$ the shell can take. The $c, c^{\dag}$ are the usual fermion operators. The operators $S_+, S_-, S_0 = \frac{1}{2}\sum_{m>0}[c^{\dag}_mc_m + c^{\dag}_{-m}c_{-m} -1]$ are pseudo-spin operators obeying the SU2 Lie algebra

\begin{equation}
  [S_-,S_+] = -2S_0~~; [S_0,S_{\pm}]= \pm S_{\pm}
  \label{comms}
\end{equation}
Because of the spin algebra, the model is readily solved leading to the eigenvalues of $H$

\begin{equation}
  E(N,s) = -\frac{G}{4}(s^2 - 2s(\Omega + 1) +2N(\Omega+ 1) - N^2)
  \label{eigenvals}
\end{equation}
with $N$ the particle number and $s$ the so-called seniority which counts the unpaired particles. Then, we have

\[ s = 0, 2, 4, ...N ~~~~~{\rm for} ~~~~ N ~~~~{\rm even}\]
\[ s = 1, 3, 5, ...N ~~~~~{\rm for} ~~~~ N ~~~~{\rm odd} \]
The unnormalized ground state wave function for $N$ even is given by

\begin{equation}
  |\Omega/2, S_0\rangle = |\Omega/2,(N-\Omega)/2\rangle \propto S_+^{N/2}|vac\rangle
  \label{gswf}
  \end{equation}
For $N$ odd, that is for a $s=1$ state the ground state wave function is given by

\begin{equation}
  |(\Omega -1)/2, S_0\rangle \propto S_+^{(N-1)/2}c^{\dag}_m|vac \rangle
  \label{gs-odd}
\end{equation}
and so forth for the $s=2$ wave function, etc. All this is well described in standard textbooks of nuclear physics, see, e.g. \cite{RS}.\\
Additionally there exists the relation

\begin{equation}
  -2S_0|\Omega/2, S_0\rangle = (\Omega - N)|\Omega/2, S_0\rangle  ~~{\rm for}~~ N<\Omega
  \label{S0}
\end{equation}
which  will be very useful below. We also will consider the chemical potentials $\mu_{\pm}$ obtained from adding or removing a particle

\begin{eqnarray}
  \mu_+&=& E(N+1,s=1)-E(N,s=0)\nonumber\\ &=& G \frac{N}{2}\nonumber\\
  \mu_- &=& E(N,s=0)-E(N-1,s=1)\nonumber\\&=& -G\frac{1}{2}[2(\Omega+1)-N]
  \label{chem-pot}
  \end{eqnarray}

\section{Dyson equation approach}

In the case of the seniority model it is of great help that the exact solution as given above, can easily be found analytically from angular momentum algebra. In other single shell models, this may not be the case. It may, therefore, be useful to give methods which solve the problem without the recourse to angular momentum algebra. We will show this here with the use of Dyson's equation and the equation of motion (EOM) method for a many-fermion system. Let us start introducing the following generalized single particle operators

\begin{eqnarray}
  q^{\dag}_{\alpha , m} &=& u^{\alpha}c^{\dag}_m + v^{\alpha}S_+c_{-m}\nonumber\\
  q_{\alpha,m}&=& u^{\alpha}c_m + v^{\alpha}c^{\dag}_{-m}S_-
  \label{add-op}
\end{eqnarray}
It can be realized that the destruction operator $q_{\alpha, m}$ kills, with certain relation between the $u, v$ amplitudes, the vacuum (\ref{gswf}) exactly, i.e., $q_{\alpha, m}|\Omega/2, S_0\rangle = 0 $. We will give more details on this later. In order to find the amplitudes $u,v$, we, as usual \cite{SCRPA} minimize the following energy weighted sum rule

\begin{equation}
  E_{\alpha} = \frac{1}{2}\frac{\langle \{q_{\alpha, m},[H,q^{\dag}_{\alpha, m}]\}\rangle}{\langle \{q_{\alpha, m},q^{\dag}_{\alpha, m}\}\rangle }
  \label{sun-rule}
\end{equation}
where $\{..,..\}$ is an anti-commutator and $\langle ... \rangle$ stands for expectation value with the ground state wave function (\ref{gswf}). The minimization yields the following eigenvalue equation

\begin{equation}
  \begin{pmatrix}a&b\\c&d\end{pmatrix}
      \begin{pmatrix}u^{\alpha}\\v^{\alpha} \end{pmatrix}
      = E_{\alpha} \begin{pmatrix}n_{11}&n_{12}\\n_{21}&n_{22}\end{pmatrix}  \begin{pmatrix}u^{\alpha}\\v^{\alpha} \end{pmatrix}
      \label{eigen-eq}
      \end{equation}
with the expressions for the various matrix elements

\begin{eqnarray}
  a &=& \langle \{c_m,[H,c^{\dag}_m]\}\rangle~;\nonumber\\
  b &=& \langle \{c_m,[H,S_+c_{-m}]\}\rangle~; ~~\nonumber\\
  c&=&  \langle \{c^{\dag}_{-m}S_-,[H,c^{\dag}_m]\}\rangle~;\nonumber\\
  d&=&\langle \{c^{\dag}_{-m}S_-,[H,S_+c_{-m}]\}\rangle
 \end{eqnarray}
and

\begin{eqnarray}
 n_{11} &=& \langle \{c_m,c^{\dag}_m\}\rangle~;~~ n_{12} = \langle \{c_m,S_+
 c_{-m}\}\rangle~; ~~ n_{21}=n_{12}~;\nonumber\\
 n_{22} &=& \langle \{c^{\dag}_{-m}S_-,S_+c_{-m}\}\rangle
 \end{eqnarray}
With (\ref{S0}) those matrix elements can be simplified. For the elements of the norm, we obtain

\begin{eqnarray}
  n_{11}&=&1;~~n_{12}=n_{21} = \frac{N}{2\Omega}\nonumber\\
n_{22}&=& (\Omega - N + 1)\frac{N}{2\Omega} + \langle S_+S_-\rangle
\label{norms}
\end{eqnarray}
where we used the fact that all the elements are independent of the index $m$, so that we can sum over $m$ and divide by $\Omega$. For example $\langle c^{\dag}_mc_m\rangle =\frac{1}{\Omega} \sum_m \langle c^{\dag}_mc_m\rangle = N/(2\Omega)$.
For the solution of the eigenvalue equation (\ref{eigen-eq}), it is convenient to absorb the norm into the eigenvector

\begin{equation}
  \begin{pmatrix}\tilde a&\tilde b\\\tilde c&\tilde d\end{pmatrix}
      \begin{pmatrix}\tilde u^{\alpha}\\\tilde v^{\alpha} \end{pmatrix}
      = E_{\alpha}  \begin{pmatrix}\tilde u^{\alpha}\\\tilde v^{\alpha} \end{pmatrix}
      \label{eigen-eq2}
      \end{equation}
The matrix elements can then be given in the following form

\[ \Sigma^{(0)} \equiv \begin{pmatrix}\tilde a&\tilde b\\\tilde c&\tilde d\end{pmatrix}=
  \begin{pmatrix}a&b\\c&d\end{pmatrix}\begin{pmatrix}n_{11}&n_{12}\\n_{21}&n_{22}\end{pmatrix}^{-1} \]
with

\begin{eqnarray}
 \tilde a &=& 0~; ~~ \tilde b/G = \tilde c/G = n_{12}~;\nonumber\\
 \tilde d/G &=& (\Omega - N + 1)^2\frac{N}{2\Omega}\nonumber\\
 &-& \bigg [ \frac{N-2}{2\Omega} - (\Omega -N + 2)\nonumber\\
   &+& \frac{\Omega - N + 1}{\omega}\bigg ]\langle S_+S_-\rangle
 \end{eqnarray}

The matrix $\Sigma^{(0)}$ is not symmetric. It is possible to use a procedure with symmetric matrices in diagonalising the norm matrix and dividing the matrices left and right by the square root of its eigenvalues. However, we will not opt for this possibility here.\\    
It can be demonstrated that the two eigenvalues $E_{\alpha}$ agree with the exact values and also that the amplitudes $\tilde u, \tilde v$ are related to the various correlation functions. However, to show this, we find it more convenient to switch to the equivalent Green's function formalism. We introduce



    \begin{equation}
      {\mathcal G}^{t-t'} = \begin{pmatrix}G_{11}&G_{12}\\G_{21}&G_{22}\end{pmatrix}
      \label{GF}
    \end{equation}
    with in standard notation

    \begin{eqnarray}
     G_{11}&=&-i\langle Tc_m(t)c^{\dag}_m(t')\rangle~;\nonumber\\
    G_{12} &=&-i\langle Tc_m(t)(S_+c_{-m})_{t'}\rangle ~;\nonumber\\
     G_{21}&=&-i\langle T(c^{\dag}_{-m}S_-)_tc^{\dag}_m(t')\rangle~;\nonumber\\
     G_{22}&=&-i\langle T(c^{\dag}_{-m}S_-)_t(S_+c_{-m})_{t'}\rangle
     \end{eqnarray}
    The integral equation (in time), equivalent to the eigenvalue equation (\ref{eigen-eq2}) for the GF is then

    \begin{equation}
      i{\partial_t} {\mathcal G} = {\mathcal N}+\Sigma^{(0)}{\mathcal G}
      \label{GF-eq}
    \end{equation}
    The norm ${\mathcal N}$ is the one which appears on the r.h.s. of eq.(\ref{eigen-eq}), that is ${\mathcal N}_{ij} \equiv n_{ij}$ .
    The spectral representation of the GF is given by (passing by Fourier transformation from time to frequency $\omega$)

    \begin{equation}
     {\mathcal G}^{\omega}= \frac{ \begin{pmatrix}u^{(1)}\\v^{(1)}\end{pmatrix}(u^{(1)} v^{(1)})}{\omega - \mu_+ + i\eta} + \frac{ \begin{pmatrix}u^{(2)}\\v^{(2)}\end{pmatrix}(u^{(2)} v^{(2)})}{\omega - \mu_- - i\eta}
      \label{spectral}
    \end{equation}
    where we realized the equivalence $E_{1/2} \equiv \mu_{\pm}$.
    We also recall that via the following sum-rule \cite{FW}, one obtains for the ground state energy

   \begin{equation}
      -i\lim_{t'\rightarrow t=0^+}i\partial_t\sum_mG_{11} = \langle H\rangle
      \label{Kolthun}
      \end{equation}
    Solving (\ref{GF-eq}) for $G_{11}$, we can read off how $\langle S_+S_-\rangle $ is related to the amplitudes $u,v$ via (\ref{Kolthun}) or directly via $G_{12}$ or $G_{21}$. Let us, therefore, establish from (\ref{GF-eq}) the equation for $G_{11}$

      \begin{equation}
        (\omega -\Sigma^{(0)}_{11})G_{11} = I_{11}  + \frac{\Sigma^{(0)}_{12}\Sigma^{(0)}_{21}}{\omega - \Sigma^{(0)}_{22}}G_{11}
          \label{Dyson-eq}
      \end{equation}
with

      \[ I_{11} =  n_{11} + \Sigma^{(0)}_{12}\frac{n_{21}}{\omega - \Sigma^{(0)}_{22}}\]
      This is the Dyson equation for the single particle GF. Its solution reads

      \[ G_{11}=\frac{n_{11}(\omega - \Sigma^{(0)}_{22}) + \Sigma^{(0)}_{12}n_{21}}{(\omega -\Sigma^{(0)}_{11})(\omega - \Sigma^{(0)}_{22}) - \Sigma^{(0)}_{12}\Sigma^{(0)}_{21}} \]
      The poles of $G_{11}$ are at

      \begin{equation}
        \mu_{\pm} = \frac{1}{2}\bigg [\Sigma^{(0)}_{11}+\Sigma^{(0)}_{22} \pm \sqrt{(\Sigma^{(0)}_{11}-\Sigma^{(0)}_{22})^2 + 4\Sigma^{(0)}_{12}\Sigma^{(0)}_{21}}\bigg ]
          \label{eigen-val}
      \end{equation}
      From where we get the spectral representation

      \begin{equation}
        G^{\omega}_{11}=\frac{\Psi_{11}^2}{\omega - \mu_+ + i\eta} + \frac{\Phi_{11}^2}{\omega - \mu_- -i\eta}
        \label{spectral-w}
      \end{equation}
      with the residua

      \begin{equation}
        \Psi_{11}^2 = (u^{(1)})^2 =\frac{n_{11}(\mu_+-\Sigma^{(0)}_{22}) + n_{12}\Sigma^{(0)}_{12}}{\mu_+ - \mu_-}
        \label{resid-for}
      \end{equation}
      and

      \begin{equation}
        \Phi_{11}^2 = (u^{(2)})^2 = -\frac{n_{11}(\mu_--\Sigma^{(0)}_{22}) + n_{12}\Sigma^{(0)}_{12}}{\mu_+ - \mu_-}
        \label{resid-back}
      \end{equation}
      Of course the completeness $\Psi_{11}^2 + \Phi_{11}^2 = 1$ is fullfilled.
      Via the sum-rule (\ref{Kolthun}), we then find the identity

      \begin{equation}
        \langle H\rangle = \mu_-\sum_m \Phi_{11}^2 =  \mu_-\Omega \Phi_{11}^2
      \end{equation}
      We can check that the equation for $\langle S_+S_-\rangle $ to which this equation leads is fullfilled by the exact expression

     \begin{equation}
       \langle S_+S_-\rangle = \frac{N}{4}(2\Omega - N +2)
       \label{S+S-}
       \end{equation}
      The chemical potentials are given by

      \begin{equation}
        \mu_{\pm} = \frac{1}{2}\bigg [\Sigma_{22} \pm \sqrt{\Sigma_{22}^2 + 4\Sigma_{12}\Sigma_{21}}\bigg ]
        \label{chem-pm}
      \end{equation}
      The expressions for the $\Sigma_{ij}$ in terms of $\Omega$ and $N$ can be found in the Appendix. They are rather complicated but it can be demonstrated that they lead to chemical potentials which are identical to the ones in Eq.(\ref{chem-pot}).\\
      An important quantity which can be calculated from the single particle GF are the occupation numbers $n_m= \langle c^{\dag}_mc_m\rangle = -i\lim_{t' \rightarrow t=0^+}G_{11}^{t-t'}$. They can be obtained from a solution of the Dyson equation what constitutes a self-consistency problem: the occupation numbers which go into the solution of the Dyson equation should be the same as the ones we get out from the GF. Since in our problem the single particle GF is exact, no wonder that this self-consistency problem leads to the exact occupancies. Indeed we can analytically verify that the hole occupancy (\ref{resid-back}) which is a function of the $n_m$ fullfills the equality

        \[ \Phi_{11}^2 = \frac{N}{2\Omega} = n_m \]
        Those occupancies are by the way the same as one gets from the BCS solution of the seniority model \cite{RS}. They are independent of the azimuthal quantum number $m$ and, thus, take fractional values determined by the ratio of the particle number $N$ and the total degeneracy of the shell $2\Omega$.
      
\noindent
      One also can check that the killing condition is fullfilled.

\section{Pair addition and pair removal modes}

It is interesting that the same game can be repeated for the pair-addition and pair removal modes. The extended RPA operators are

\begin{equation}
  A^{\dag} = X_1 S_+ + X_2S_+S_0
  \label{A+}
\end{equation}
for the pair addition, and

\begin{equation}
  R^{\dag} = Z_1S_- + Z_2S_0S_-
  \label{R+}
\end{equation}
for the removal.\\

Again one sees that it is possible to fullfill the killing conditions

\[ A|0\rangle = R|0\rangle = 0 \]
Of course the addition and removal modes should be properly normalized

\[ \langle [A,A^{\dag}]\rangle = 1~;~~ \langle [R,R^{\dag}]\rangle = -1 \]
By simple manipulations with back and forward amlitudes we can invert (\ref{A+}) and (\ref{R+}) and derive the following expressions for $S_+$ and $S_+S_0$

\begin{equation}
  S_+ = \frac{Z_2A^{\dag}-X_2R}{Z_2X_1-X_2Z_1}~; S_+S_0 = \frac{Z_1A^{\dag}-X_1R}{Z_1X_2-X_1Z_2}
  \label{inv}
\end{equation}
For example this allows to obtain the realtion

\[\langle S_+S_-\rangle = \frac{X_2^2}{(Z_2X_1-X_2Z_1)^2} \]
  From the EOM we obtain the eigenvalue equation for addition and removal modes

  \begin{eqnarray}
    eX_1 + fX_2 &=& \omega_a (N_{11}X_1 + N_{12}X_2)\nonumber\\
    fX_1 + gX_2 &=& \omega_a (N_{21}X_1 +N_{22}X_2)\nonumber\\
    &~&\nonumber\\
    eZ_1 + fZ_2 &=& -\omega_r (N_{11}Z_1 + N_{12}Z_2)\nonumber\\
    fZ_1 + gZ_2 &=& -\omega_r (N_{21}Z_1 + N_{22}Z_2)
  \end{eqnarray}
  with

  \begin{eqnarray}
    e&=&\langle [S_-,[H,S_+]]\rangle \nonumber\\
    f&=&\langle [S_0S_-,[H,S_+]]\rangle \nonumber\\
    g&=&\langle [S_0S_-,[H,S_+S_0]]\rangle
  \end{eqnarray}
  and

  \begin{eqnarray}
    N_{11}&=&\langle [S_-,S_+]\rangle \nonumber\\
    N_{12}&=&N_{21}= \langle [S_-,S_+S_0]\rangle \nonumber\\
    N_{22}&=&\langle [S_0S_-,S_+S_0]\rangle
    \end{eqnarray}

With these EOMs and proceding in a similar way as with the Dyson equation one also obtains the exact solutions for the s=0 ground state and addition and removal energies. It is also possible to introduce a corresponding system of GFs, establish a Dyson-Bethe-Salpeter equation (Dyson-BSE), see \cite{Toulouse}, and solve the problem in this way. So, we got the exact solutions for the $s=0$ ground states of the $N,N\pm2$ systems and for the $s=1$ systems with $N\pm 1$. Higher seniority states need further elaboration.\\

\section{Outlook and conclusions}

It is very interesting that the above procedure may give way how to deal with the general more level case. We know that a general coupled cluster doubles wave function $|CCD\rangle $ can be killed by an extended RPA operator, see \cite{SCRPA}. To be short but without loss of generality, we take the two level Lipkin model as example. For the Lipkin model, see, e.g., \cite{RS}, the extra term of the generalized killing operator is contained in

\begin{equation}
  Q^+ = XJ_+ - YJ_- +\eta YJ_-J_0
  \label{Q-op}
  \end{equation}
with, as we know \cite{SCRPA}

\[ Q|CCD\rangle = 0 \]
and $|CCD\rangle = e^{zJ_+J_+}|HF\rangle$ with amplitudes fullfilling certain inter-relations.
The $J_-J_0$ term is analogous to the $S_0S_-$ term in the pair operators. The $J_{\pm},J_0$ operators are quasi-spin operators as in the seniority model here.
The RPA operator in (\ref{Q-op}) has three terms because the Lipkin model is a two level model and, thus, the killer has essentially already the same structure as the killer in a realistic case \cite{SCRPA}. In the past we have replaced in (\ref{Q-op}) $J_0$ by its expectation value $\langle J_0\rangle$ and established self-consistent RPA (SCRPA) equations \cite{SCRPA}. The $J_-J_0$ operator may, however, be better taken care of. In analogy to (\ref{GF}) we can construct a 3$\times$3 GF ${\mathcal G} = G_{ij}$ with $i,j = 1,2,3$ and establish a Dyson equation analogous to (\ref{GF-eq}, \ref{Dyson-eq}) but the single particle GF's replaced by the analogons to the pair modes. In the Lipkin model the correlated ground state is not an eigenstate of $J_0$ as is the case in the seniority model with $S_0$. However in the relevant terms of $\Sigma_{\rm Lipkin}$ the $J_0$ operators can be moved to the right (or left) until it touches the ground state and, then taken out of the correlation function with its expectation value. This may be a quite reasonable approximation, probably a better one than to replace $J_0$ by its mean value in (\ref{Q-op}). Separate expressions for $\langle J_0\rangle $ can be established as the one for $G_{11}$ here, or similar ones \cite{mohsen}. This will give rise to a coupled system of a single fermion Dyson equation and a Dyson-Bethe-Salpeter equation \cite{Toulouse}. We are then in an analogous situation as with the present seniority model and a fully self consistent solution can  also be given for the Lipkin model or even for more realistic problems.
This shall be a task for the future.\\
In conclusion, we have found with the equation of motion method a procedure which solves the single level pairing (the seniority) model exactly. This may be useful for the solution of other single level models which cannot be solved trivially by angular momentum algebra. The applied many body technique is rather non-trivial and it may be generalizable to more realistic many level problems. Work in this direction is in progress.

\section{Acknowledgements}

I want to mention that the formalism of the pair modes is given in old unpublished notes of my former Diploma student Georg Wagner (in the private now).
I also want to thank M. Jemai for help with some algebraic manipulations and for a careful reading of the manuscript. Discussions and interest by Thierry Champel have been appreciated. Thanks also go to Jorge Dukelsky for helpful remarks.

\section{Appendix}

The matrix elements of the self energy $\Sigma^{(0)}$ are given by

\begin{eqnarray}
  \Sigma_{22} &=& -\frac{G}{D}[C-n_{22}n_{12}]\nonumber\\
  \Sigma_{12} &=& -G\nonumber\\
  \Sigma_{21} &=& -\frac{G}{D}\bigg \{ \bigg [\frac{(\Omega-N+1)N}{2\Omega}+\langle S_+S_-\rangle\bigg ]^2\nonumber\\
 ~~~ &-&C\frac{N}{2\Omega}\bigg \}
\end{eqnarray}
with the following ingredients (for $n_{ij}$ and $\langle S_+S_-\rangle$ see (\ref{norms}) and (\ref{S+S-}), respectively)


\begin{eqnarray}
  C &=& (\Omega - N + 1)^2\frac{N}{2\Omega}\nonumber\\
  &-& \bigg [ \frac{N-2}{2\Omega} - (\Omega -N + 2)\nonumber\\
    &+& \frac{\Omega - N + 1}{\Omega}\bigg ]\langle S_+S_-\rangle 
\end{eqnarray}

\begin{equation}
  D = \frac{(\Omega - N + 1)N}{2\Omega} -\bigg (\frac{N}{2\Omega} \bigg )^2 + \langle S_+S_-\rangle
  \end{equation}
With these expressions one can show that

\begin{eqnarray}
  D\Sigma_{22}&=& -G(\Omega - N + 1)D\nonumber\\
  D\Sigma_{21} &=& -G\frac{N}{4}(2\Omega - N + 2)D
\end{eqnarray}
that is, the numerator of $\Sigma_{ij}$ factorizes.
With this it is easy to show that (\ref{chem-pm}) is equivalent to (\ref{chem-pot}).










\begin{thebibliography}{99}
  \bibitem{RS}
    P. Ring, P. Schuck, {\it The nuclear Many-Body Problem}, Springer 1980.
    
  \bibitem{Racah}
  G.  Racah, Phys. Rev. 63, 367 (1943).

  \bibitem{Kerman}
A. K. Kerman, Ann. Phys. (New York) 12, 300 (1961).


\bibitem{SCRPA}
  M. Jemai, D. S.  Delion, P. Schuck, Phys. Rev. C 88, 044004 (2013).

\bibitem{FW}
A. L. Fetter, J. D. Walecka, {\it Quantum Theory of Many-Particle Systems}, Dover Publications, Inc., Mineola, New York.

\bibitem{Toulouse}
  V. Olevano, J. Toulouse, P. Schuck, J. Chem. Phys. 150, 084112 (2019).


  \bibitem{mohsen}
  M. Jemai, P. Schuck, Phys. Rev. C 100, 034311 (2019).


  
\end{thebibliography}
\end{document}